\documentclass[fleqn,usenatbib]{mnras}

\usepackage{hyperref}
\usepackage{newtxtext,newtxmath}

\usepackage[T1]{fontenc}
\usepackage{ae,aecompl}

\usepackage{graphicx}	
\usepackage{amsmath}	
\usepackage{amssymb}	
\usepackage{subfig}
\usepackage[wby]{callouts}

\graphicspath{ {images/} }

\def\Fig{\mbox{Fig.~}}

\def\Tab{\mbox{Table~}}
\def\Sec{\mbox{Sect.~}}
\def\Secs{\mbox{Sections~}}

\def\lsim{\mathrel{\rlap{\lower3.5pt\hbox{\hskip0.5pt$\sim$}}
    \raise0.5pt\hbox{$<$}}}
\def\gsim{~\rlap{$>$}{\lower 1.0ex\hbox{$\sim$}}}

\title[Testing ConvNet lens-finders]{Testing Convolutional Neural Networks for finding strong gravitational lenses in KiDS}

\author[C.~E.~Petrillo et al.]{C.~E.~Petrillo$^{1}$\thanks{E-mail: petrillo@astro.rug.nl},  C.~Tortora$^{1}$, S.~Chatterjee$^{1}$, G.~Vernardos$^{1}$, L.~V.~E.~Koopmans$^{1}$, \and G.~Verdoes~Kleijn$^{1}$, N.~R.~Napolitano$^{2}$, G.~Covone$^{3,4}$, L. S. Kelvin$^{5}$, A. M. Hopkins$^{6}$\vspace{0.2cm}\\ 
$^{1}$Kapteyn Astronomical Institute, University of Groningen, Postbus 800, 9700 AV, Groningen, The Netherlands\\
$^{2}$INAF - Osservatorio Astronomico di Capodimonte, Salita Moiariello, 16, 80131 Napoli, Italy\\
$^{3}$Dipartimento di Fisica ``E. Pancini", Universit\`a di Napoli Federico II, Compl. Univ. Monte S. Angelo, 80126 Napoli, Italy\\
$^{4}$INFN, Sezione di Napoli, Compl. Univ. Monte S. Angelo, 80126 Napoli, Italy\\
$^{5}$Astrophysics Research Institute, Liverpool John Moores University, IC2, Liverpool Science Park, 146 Brownlow Hill, Liverpool, L3 5RF\\
$^{6}$Australian Astronomical Optics, Macquarie University, 105 Delhi Rd, North Ryde, NSW 2113, Australia
}

\date{Accepted XXX. Received YYY; in original form ZZZ}

\pubyear{2015}

\begin{document}
\label{firstpage}
\pagerange{\pageref{firstpage}--\pageref{lastpage}}
\maketitle

\begin{abstract}
Convolutional Neural Networks (ConvNets) are one of the most promising methods for identifying strong gravitational lens candidates in survey data. 
We present two ConvNet lens-finders which we have trained with a dataset composed of real galaxies from the Kilo Degree Survey (KiDS) and simulated lensed sources. One ConvNet is trained with single \textit{r}-band galaxy images, hence basing the classification mostly on the morphology. While the other ConvNet is trained on \textit{g-r-i} composite images, relying mostly on colours and morphology. We have tested the ConvNet lens-finders on a sample of 21789 Luminous Red Galaxies (LRGs) selected from KiDS and we have analyzed and compared the results with our previous ConvNet lens-finder on the same sample. 
The new lens-finders achieve a higher accuracy and completeness in identifying gravitational lens candidates, especially the single-band ConvNet. Our analysis indicates that this is mainly due to improved simulations of the lensed sources.
In particular, the single-band ConvNet can select a sample of lens candidates with $\sim40\%$ purity, retrieving 3 out of 4 of the confirmed gravitational lenses in the LRG sample. 
With this particular setup and limited human intervention, it will be possible to retrieve, in future surveys such as Euclid, a sample of lenses exceeding in size the total number of currently known gravitational lenses.
\end{abstract}




\section{Introduction}
Strong gravitational lensing is a phenomenon that originates when light rays propagating from a background source galaxy are deflected, on their way towards the observer, by the gravitational field of a foreground galaxy, creating multiple images, arcs and/or rings around the foreground galaxy.
Strong gravitational lensing is a unique probe for studying the (dark) matter distribution of galaxies and providing cosmological constraints.
E.g., gravitational lenses have been used to measure the Hubble constant through time delays of lensed quasar images (e.g., \citealt{suyu2010,Suyu2017,bonvin2016}) and to constrain the dark energy equation of state (e.g., \citealt{Biesiada+10,Collet2014de,Cao+12,Cao2015de}).
Gravitational lensing also allows measuring the fraction of dark matter in the central regions of galaxies (\citealt{gavazzi2007,jiang2007,covone2009,grillo2010,cardone2009,Cardone2010,Auger+10_SLACSX,tortora2010central,more2011,ruff2011,sonnenfeld2015sl2s})
and to constrain the slope of the inner mass density profile (e.g., \citealt{Treu2002MNRAS,Treu2002,koopmans2006,koopmans2003,Moore2008,barnabe2009,koopmans2009,cao2016limits,Mukherjee2018}). 
Moreover, studying gravitational lenses can constrain the stellar initial mass function (e.g., \citealt{treu2010, ferreras2010, Spiniello:2011p8239, brewer2012swells,sonnenfeld2015sl2s,posacki2015stellar,Spiniello+15_IMF_vs_density,moller2007strong,Sonnenfeld2018imf}).
Strong lensing also works as a ``cosmic telescope", producing a magnified view of background objects otherwise not observable (e.g., \citealt{impellizzeri2008,swinbank2009,richard2011,deane2013,treu2015grism,Mason2016,Salmon2017,Kelly2017}).
Discovering new gravitational lenses allows placement of more precise constraints on the above-mentioned quantities (see e.g., \citealt{Vegetti2009,Barnab2011,Li2016}).
For a comprehensive review of the scientific applications of strong lensing see, e.g., \cite{schneider1992gravitational}, \cite{Schneider06_SAASFEE33} and \cite{Treu2010rev}.

Originally, gravitational lenses were found serendipitously in astronomical surveys, while currently they are considered as an important class of objects to systematically search in large sky surveys.
The most successful campaign aiming at building a homogeneous dataset of strong gravitational lenses was the Sloan Lens ACS Survey (SLACS; \citealt{SLACS2008}) with more than 100 observed lenses that were identified by analyzing spectra from the Sloan Digital Sky Survey (SDSS; \citealt{SDSS}) that exhibited the imprint of two galaxies at two different redshifts.
On-going optical surveys such as the Hyper Suprime-Cam survey (HSC; \citealt{HSCmiyazaki2012hyper}), the Kilo Degree Survey (KiDS; \citealt{deJong+15_KiDS_paperI}) and the Dark Energy Survey (DES; \citealt{DES}) are expected to provide in the coming years thousands of new lenses (see \citealt{collet2015} and \citealt{Petrillo2017}) and have already provided new lens candidates \citep{Petrillo2017,Diehl2017,Sonnenfeld2018}.
The future is bright 
also in the sub-millimeter wavelength, where Herschel \citep{negrello2010} and the South Pole Telescope (\citealt{carlstrom2011}), coupled with the Atacama Large Millimeter/sub-millimeter Array, are providing
several hundreds of new lens candidates \citep{vieira2013dusty,negrello2017herschel}.
However, it is the next decade that holds a treasure trove of new gravitational lenses. It has been estimated that samples of $\sim 10^5$ strong lenses \citep{oguri2010gravitationally,pawase2012,collet2015,McKean2015} will be observed by Euclid \citep{Laureijs:2011wi}, the Large Synoptic Survey Telescope (LSST; \citealt{abell2009}) and the Square Kilometer Array\footnote{\href{https://www.skatelescope.org/}{\tt https://www.skatelescope.org/}}.

The huge number of possible new candidates, together with the difficulty of identifying them in the enormous volume of survey data, drives the growing effort in developing automatic lens-finders. 
Most are based on the identification of arc-like features (e.g., \citealt{Lenzen2004,Horesh2005,Alard2006,Estrada2007,Seidel2007,Kubo2008,More2012, Maturi2014}).  
Other approaches, such as described by \cite{Gavazzi2014} and \cite{Joseph2014}, focus on the analysis of the residuals after subtracting the candidate lens galaxies from the astronomical images. Both methods have been used to find lens candidates in the Canada-France-Hawaii Telescope
Legacy Survey (CFHTLS\footnote{\href{http://www.cfht.hawaii.edu/Science/CFHLS/}{\tt http://www.cfht.hawaii.edu/Science/CFHLS/}}) by \cite{Sonnenfeld2013,Gavazzi2014,Paraficz2016}. 
Instead, the algorithm developed by \cite{Chan2015} is specialized in identifying lensed quasars and together with the algorithm \textsc{YattaLens} \citep{Sonnenfeld2018} has been applied to find lens candidates in HSC \citep{Sonnenfeld2018}.
Another approach, as in \cite{Brault2015}, is modeling the probability that the targets are actual lenses. \cite{Stapelberg2017} applied the same strategy to clusters and groups of galaxies. Gravitational lenses have been identified also with citizen-science experiment approaches with the Space Warps project \citep{Marshall2016,more2016} where 
non professional volunteers can classify galaxy images with the help of a web applet\footnote{\href{https://spacewarps.org/}{\tt https://spacewarps.org/}}.
Most recently, \cite{Petrillo2017} and \cite{Jacobs2017} have used Convolutional Neural Network (ConvNets) for finding lens candidates in KiDS and CFHTLS, respectively. Finally, \cite{Hartley2017} have used a technique based on Support Vector Machines (SVMs) and applied it to KiDS. Instead \cite{Spiniello2018} focused on the search of lensed quasars in KiDS using 3 different morphology based methods.

In this paper we present and test our latest ConvNet lens-finders, improving on the work of \cite{Petrillo2017}. ConvNets (\citealt{fukushima1980neocognitron,lecun1998gradient}) are the state of the art and often the standard choice among machine learning algorithms for pattern recognition in digital images. The winners of the ImageNet Large Scale Visual Recognition Competition (ILSVRC; \citealt{ILSVRC15}; the most important image classification competition) in recent years have all been groups utilizing ConvNets. The advantage of the latter method with respect to other pattern recognition algorithms is that the features are not hand-crafted but are themselves extracted automatically during the training procedure, thus the algorithm decides which features are most representative for classifying the images.
The theoretical basis of ConvNets was developed in the 1980s and the 1990s. However only recently ConvNets have started to outperform other algorithms thanks to the advent of large labeled datasets, improved algorithms and faster training, especially on Graphics Processing Units (GPUs). The interested reader is referred to the Appendix for a brief introduction on ConvNets and to the reviews by \cite{schmidhuber2015deep}, \cite{lecun2015deep} and \cite{guo2016deep} for a more detailed introduction.

ConvNets have been used recently in many astronomical problems, e.g., galaxy morphology classification \citep{dieleman2015rotation,Huertas2015}, estimation of photometric redshifts \citep{Hoyle2016,D'Isanto2018}, spectra classification \citep{Hala2014,Tao2018}, identifying exoplanets \citep{Shallue2018}, transient detection \citep{Cabrera-Vives2017}, galaxy surface brightness estimation \citep{Tuccillo2018}, strong lensing parameters estimation \citep{Hezaveh2017} and star/galaxy separation \citep{Kim2016}. 

More importantly, \cite{Metcalf2018} presented the results of a large international {\sl challenge} in which various methods of identifying simulated gravitational lenses were tested blindly. This challenge, the first of a series, sets out to prepare the community for finding lenses in the data of ESA's Euclid mission. Its large data volume requires fast and efficient algorithms to identify strong gravitational lenses. However, the methods were also tested on simulated KiDS data. 
ConvNets and Support Vector Machines (SVMs) were recognized to be the most promising methods, among many different methods tested in the challenge.

The ConvNet lens-finders presented in this paper will be applied on $\sim$ 900 sq. deg of the KiDS survey in a forthcoming paper with the purpose of starting a systematic census of strong lenses named ``LinKS" (Lenses in KiDS Survey).
The paper is organized as follows. 
In \Sec\ref{SECtraining}, we illustrate our lens-finding ConvNet-based algorithms and how the training dataset is built. 
In \Sec\ref{SECanalysis}, we evaluate the performances of the ConvNets.
In \Sec\ref{SECrealdata}, we apply the lens-finders to $\sim22000$ extracted from $\sim255$ square degrees of KiDS for testing the algorithms on real data. 
Finally, in \Sec\ref{SECdiscussion}, we provide a summary and the main conclusions of this work.

\section{training the CONVNETS to find strong lenses}\label{SECtraining}
A Convolutional Neural Network (ConvNet) can be seen as a sequence of non-linear functions, called layers, that create, starting from an input image, a series of increasingly abstract representations of the input called feature maps. The final layer of the ConvNet converts the input feature maps into a set of numbers that represent the outcome of the classification. Hence a ConvNet maps an image onto a single or few numbers.
In our case the output is a single number, denoted by $p$, which can vary between 0 and 1, and it is related to the probability that the input image is a lens (see \citealt{saerens2002} for a detailed discussion).
The parameters of the non-linear functions are obtained during the so called training phase where labeled images are fed to the ConvNet. In more detail, the parameters are derived by minimizing a loss function that expresses the difference between the label values of the images (1 for lenses, 0 for non-strong-lensing systems) and the output $p$ of the ConvNet. 
Although in \cite{Petrillo2017} we have used a similar set-up, the aim of this work is to improve the performance of our previous lens-finder.

Currently we use a ConvNet with a \textit{ResNet}-type architecture that has 18 layers, exactly as described in \cite{he2015deep}. ResNet-type architectures are often the preferred choice in image classification tasks  because of their faster convergence and higher classification accuracy with respect to other architectures. Moreover, ResNet architectures have already been tested successfully on identifying simulated lenses and they have proven to be one of the best architecture for this task \citep{Schaefer2017,lanusse2018,Metcalf2018}.
We train two different ConvNets with the same architecture except that one takes in input RGB-images composed with \textsc{HumVI}\footnote{{\tt https://github.com/drphilmarshall/HumVI}} \citep{Marshall2016}, while the other takes single \textit{r}-band images as input. We choose the \textit{r}-band as single-band input because the KiDS observing strategy reserves the best seeing conditions for this band (which is used for the weak lensing studies; \citealt{Kuijken2015,Hildebrandt2017}).
The technical details of the ConvNet and of the training procedure are described in Appendix A together with a brief introduction on ConvNets.

To produce the data used to train and validate the ConvNets, we adopt a hybrid approach similarly as done in \cite{Petrillo2017,Jacobs2017,Pourrahmani2018}, creating mock images of strong gravitational lenses using images of real galaxies from KiDS 
and super-imposing simulated lensed images.
We adopt this approach because we do not have a sample of genuine KiDS lenses large enough to train a ConvNet (usually of the order of $10^6$).

\subsection{Data}

In this section we describe the dataset used to train the ConvNets, which is composed of real KiDS galaxies and simulated lensed sources.

\subsubsection{Luminous Red Galaxies}\label{sec:LRGs}
We use the sample of Luminous Red Galaxies (LRGs; \citealt{Eisenstein2001}) presented in \cite{Petrillo2017}. We choose to focus on massive early-type galaxies, because it has been estimated that these galaxies form $\sim80\%$ of the lens-galaxy population \citep{turner1984statistics,fukugita1992statistical,kochanek1996flat,chae2003cosmic,oguri2006image,moller2007strong}. Spiral galaxies form the other $\sim$20\% but are much harder to identify.
This training sample of LRGs is a random subset of 6554 galaxies from a parent sample of 21789 selected from 255 square degrees of KiDS DR3 \citep{deJong+17_KiDS_DR3} with the following criteria (see \citealt{Petrillo2017} for more details): 

\noindent (i) The low-$z$ ($z<0.4$) LRG colour-magnitude selection of \cite{Eisenstein2001}, adapted to include more sources, both fainter and bluer:
\begin{equation}
\begin{split}
&r<20 \\
&|c_{\rm{perp}}| < 0.2 \\
&r<14+c_{\rm{par}}/0.3 \\
\text{where}\\
&c_{\rm{par}}=0.7(g-r)+1.2[(r-i)-0.18)]\\
&c_{\rm{perp}}=(r-i)-(g-r)/4.0-0.18
\end{split}
\end{equation}

\noindent (ii) A source size in the \textit{r}-band larger than the average FWHM of the PSF of the respective tiles, times an empirical factor to maximize the separation between stars and galaxies.

\subsubsection{Contaminants}\label{sec:contaminants}
Moreover, we have used a set of $\sim 6000$ KiDS sources to train the ConvNets to recognize sources that would likely be incorrectly classified as lenses otherwise,  either because they can resemble lensing features or they are ``ghosts'', i.e. they are undetected, at least significantly, in the luminous red galaxies sample discussed in \Sec\ref{sec:LRGs}.
\begin{itemize}
\item $\sim$2000 sources wrongly classified as lenses in previous tests with ConvNets identified by the authors. This is done to teach the ConvNets not to replicate previous mistakes;
\item  $\sim$3000 randomly extracted KiDS sources with \textit{r}-band magnitude brighter than 21. To provide the network with general true negatives. 
\item $\sim$1000 KiDS sources visually classified as spiral galaxies from an on-going new project of GalaxyZoo (\citealt{Willett2013}, Kelvin et al., in prep.). This is done to decrease the false positives due to spiral features. To select the galaxies we used a preliminary reduced version of the GAMA-KiDS Galaxy Zoo catalogue for the GAMA09 9h region (see \citealt{GAMAdriver2011} for further details). This catalogue contains $\sim10^4$ sources out to a redshift of $z=0.15$. We select galaxies for which a large majority of people replied to the question \textit{``Is the galaxy in the centre of the image simply smooth and rounded, or does it have features?"} with \textit{``it has features"} \footnote{The actual selection is done by selecting sources from the catalogue with a value of the attribute {\tt features\_features\_frac} larger than $0.6$.}
\end{itemize}

There is a non-zero probability that among the contaminants and the LRGs described in the previous Section there are actual gravitational lenses. We can estimate that the percentage would be of the order of $10^{-2}$ among the contaminants and $\sim1\%$ among the LRGs \citep{Petrillo2017}. Thus, even if real lenses are actually in the training sample, with such a small percentage they would not contaminate the training procedure.

\subsubsection{Mock lensed-sources}\label{SECsims}
We simulate $10^6$ lensed images of $101 \times 101$ pixels, 
using the same spatial resolution of KiDS ($\sim0.2$ arcsec per pixel), corresponding to a $20 \times 20$ arcsec field of view. 
To produce more realistic lensing systems, we add more complexity both in the source and in the lens plane with respect to the simulations in \cite{Petrillo2017}. 
The distribution of the lens and source parameters that we choose for simulating the lensed images are chosen to create a wide range of realistic lensed images. They are not meant to statistically represent a real lens population, since the training set has to be populated sufficiently densely in the parameter space to allow the ConvNets to learn all the possible configurations and to recognize lenses that are rare in a real distribution (or currently even unknown). To ensure this, a more homogeneous distribution of the parameters is advantageous in order not to over-train on the most common lens configurations.

We proceed in the following way, we sample the parameters of the Singular Isothermal Ellipsoid (SIE; \citealt{KSB_SIE94}) and \cite{Sersic68} source models as listed in \Tab\ref{TABLEmocksourceslens}. The values of the lens Einstein radius and the source effective radius are drawn from a logarithmic distribution, while the remaining parameters, listed in \Tab\ref{TABLEmocksourceslens}, are drawn from a uniform distribution.
In this way our simulation sample  contains a higher fraction of smaller rings and arcs compared to \cite{Petrillo2017} for making the new ConvNets more sensitive to this kind of objects with respect to the old one.
The source positions are chosen uniformly within the radial distance of the tangential caustics plus one effective radius of the source S\'ersic profile. This leads our training set to be mostly composed of high-magnification rings, arcs, quads, folds and cusps rather than doubles \citep{schneider1992gravitational}.
To add complexity in the lensed sources, besides the S\'ersic profile, we add between 1 and 5 small circular S\'ersic blobs. The centers of these blobs are drawn from a Gaussian probability distribution function (PDF) around the main S\'ersic source. The width of the standard deviation of the PDF is the same as the effective radius of the main S\'ersic profile. The sizes of the blobs are chosen uniformly within 1-10$\%$ of the effective radius of the main S\'ersic profile. The S\'ersic indices of the blobs are drawn using the same prescription as for the main central source. The amplitudes of the blobs are also chosen from a uniform distribution in such a way that the ratio of the amplitude of an individual blob to the amplitude of the main S\'ersic profile is at most 20$\%$.

Moreover, we add Gaussian Random Field (GRF) fluctuations to the lens potential, which, to a first order approximation, make the lens sample more realistic by adding small scale substructures \citep{SAIKAT}.
The GRF realizations we added in our simulations all follow a power law power-spectrum with a fixed exponent $-6$, which is to the first order a good approximation of substructures in lens plane in the $\Lambda$CDM paradigm \citep{hezaveh}. The variances of the realizations are drawn from a logarithmic distribution between $10^{-4} - 10^{-1}$ about mean zero in the units of square of the lensing potential. This yields both structured sources and lenses that are not perfect SIE.

For each source a realistic color is simulated to create 
images in \textit{g}, \textit{r}, \textit{i}-bands. 
In order to produce realistic 3-band images we extract magnitudes from ``COSMOS" models in \textsc{Le Phare} (\citealt{Arnouts+99}; \citealt{Ilbert+06}). This library of spectra, consists of 31 models, used for COSMOS photo-z (\citealt{Ilbert+09_COSMOS}). The basic ``COSMOS'' library is composed of 8 templates for elliptical/S0 galaxies, 11 for spiral types, and 12 for galaxies with star-burst ages ranging from
0.03 to 3 Gyr, allowing us to span a wide range of galaxy types and colours. In order to simulate the typical blue arcs observed in most of the observed lenses, we choose models bluer than S0 and calculate observer-frame magnitudes in the three KiDS wavebands \textit{g}, \textit{r} and \textit{i} for model spectra redshifted up to a redshift of $z = 3$ with a $0.1$ binning.
%
%
Moreover, to populate the magnitude-space more uniformly, we perturb the three magnitudes adding to each of them a random number uniformly extracted from the range $[-0.1,0.1]$ mag. 
We also take into account dust extinction by considering a color excess $E(B-V)$, we extract it from a normal distribution with $\sigma=0.1$ and mean 0 considering only the positives values. In this way we obtain a small extinction correction in order to avoid very red lensed sources which, in the real universe, are much rarer than blue ones.
We adopt a typical extinction curve with $R_V=3.1$, using the relation $A_x = R_x \, E(B-V)$ where $x$ represents the value for the \textit{g}, \textit{r} and \textit{i} SDSS-filters that can be found in \Tab 2 of \cite{Yuan13}. 
Finally, we convolve the three images with an average KiDS--DR3 PSF for each different band: with a FWHM of $\sim0.86$ arcsec for \textit{g}, $\sim0.68$ arcsec for \textit{r} and $\sim0.81$ arcsec for \textit{i} \citep{deJong+17_KiDS_DR3}.

\begin{table}
\caption{Range of parameter values adopted for simulating the lensed sources. The parameters are drawn uniformly, except for Einstein and effective radius, as indicated. See \Sec\ref{SECsims} for further details.}
	\begin{center}
		\begin{tabular}{l l c}
Parameter              & Range & Unit \\
\hline
\multicolumn{3}{c}{Lens (SIE)}\\
\hline
Einstein radius      & 1.0 - 5.0 (log)& arcsec\\
Axis ratio           & 0.3 - 1.0  & -\\
Major-axis angle     & 0.0 - 180 & degree\\
External shear       & 0.0 - 0.05 & -\\
External-shear angle & 0.0 - 180 & degree\\
\hline
\multicolumn{3}{c}{Main source (S\'ersic)}\\
\hline
Effective radius ($R_{eff}$)     & 0.2 - 0.6 (log)& arcsec\\
Axis ratio           & 0.3 - 1.0 & -\\
Major-axis angle     & 0.0 - 180 & degree\\
S\'ersic index       & 0.5 - 5.0 & -\\
\hline
\multicolumn{3}{c}{S\'ersic blobs (1 up to 5)}\\
\hline
Effective radius     & $(1\% - 10\%) R_{eff}$ & arcsec\\
Axis ratio           & 1.0 & -\\
Major-axis angle     & 0.0 & degree\\
S\'ersic index       & 0.5 - 5.0 & -\\

\hline
		\end{tabular}
		\label{TABLEmocksourceslens}
	\end{center}
\end{table}

\captionsetup[subfigure]{labelformat=empty}
 \begin{figure*}
   \centering
   \hspace{\fill}
   \subfloat[]{\includegraphics[width=35mm]{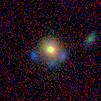}}
   \subfloat[]{\includegraphics[width=35mm]{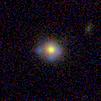}}
   \subfloat[]{\includegraphics[width=35mm]{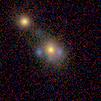}}
   \subfloat[]{\includegraphics[width=35mm]{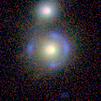}}
   \subfloat[]{\includegraphics[width=35mm]{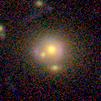}}
\caption{Examples of RGB images of simulated strong lens galaxies used to train the ConvNets. The lens galaxies are observed KiDS galaxies, while the lensed sources are simulated, as described in \Sec\ref{SECsims}.}
 \label{FIGmocklenses}
 \end{figure*}

\subsection{Creating the training set}\label{SECcreatingtrain}
The data presented above are used to build the training set which is composed of mock strong-lensing systems (labeled with a 1) and non-strong-lensing systems (labeled with a 0), i.e., objects without lensing features. In the following we outline the procedure used to build the two kinds of objects in the training set.
\\
\\
\noindent {\bf Mock strong-lensing systems}:
To create mock strong-lensing systems we carry out the following procedure: 
\\
(i) We randomly choose a mock lensed source (Sect.~\ref{SECsims}) and a LRG (Sect.~\ref{sec:LRGs});
we rescale the brightness of the simulated source to the peak brightness of the LRG in the $r$-band multiplied by a factor $\alpha$ randomly drawn from the interval $[0.02, 0.3]$. This accounts for the typical lower brightness of the lensing features with respect to the lens galaxies;
\\
(ii) we stack the LRG and the mock source for each one of the three bands;
\\
(iii) for the single-band images, we clip the negative values of the pixels to zero and performing a square-root stretch of the image to emphasize lower luminosity features. Instead, we create 3-band images with \textsc{HumVI} that operates an arcsinh stretch of the image following the \cite{Lupton2004} composition algorithm;
\\
(iv) finally, we normalize the resulting images by the galaxy peak brightness (only for single-band images).

Some examples of mock strong-lensing systems obtained in this way are shown in \Fig\ref{FIGmocklenses}.
\\
\\
\textbf{Non-strong-lensing systems}
To create the non-strong-lensing system sample we carry out the following procedure: 
\\
(i) we choose a random galaxy from either the LRG sample (with a probability of 20\%) or from the contaminant sample (80\% probability);
\\
(ii) we clip the negative values of the pixels to zero and performing a square-root stretch of the images. We create 3-band images with \textsc{HumVI};
\\
(iii) we normalize the images by the galaxy peak brightness (only for single-band images).\\

\label{sec:Dataugm}
Finally, we augment the images, which is a standard procedure in machine learning (see e.g., \citealt{Simard2003}). It is used to avoid over-fitting by expanding artificially the training set through different transformations of the images.
Before feeding the images to the ConvNets, we apply the following transformations:

\noindent
(i) a random rotation between 0 and $2\pi$;
\\
(ii) a random shift in both $x$ and $y$ direction between -4 and +4 pixels; 
\\
(iii) a $50\%$ probability of horizontally flipping the image; 
\\
(iv) a rescaling with a scale factor sampled log-uniformly between $1$ and $1.1$.

\noindent
All transformations are applied to both the mock strong-lensing systems and the non-strong-lensing systems.
The final set of inputs of the ConvNets are postage stamps of 101 times 101 pixels which correspond to $\sim 20 \times 20$ arcsec. 
The images are produced in real-time during the training phase. For more details on the training phase see Appendix A.

\section{Analysis}\label{SECanalysis}
After the training is completed, the ConvNets must be tested in order to assess whether the training was successful.
In this section we define the metric for evaluating the results and evaluate the performances of the ConvNets on a dataset composed by non-lenses and mock lenses in comparison to \cite{Petrillo2017}.

\begin{figure}
\begin{center}
 {\includegraphics[width=90mm]{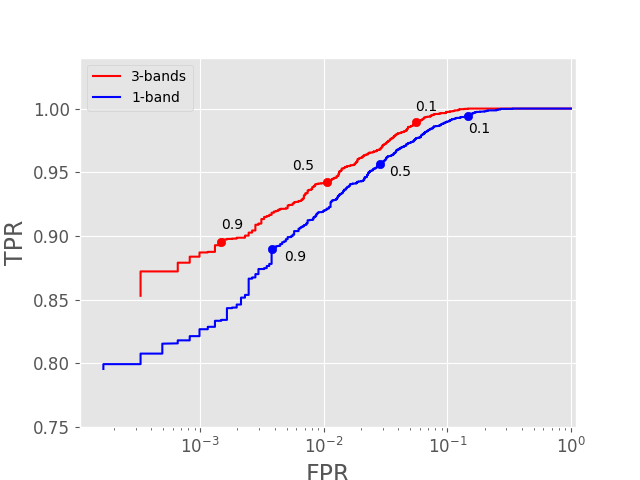}}
\caption{ROC curves for the 1-band (blue) and 3-band (red) ConvNet. Each point of the curves is the true positive rate vs. false positive rate for different values of threshold for $p$ (decreasing from left to right; some values are shown on the curve for reference).}
\label{ROC curves}
\end{center}
\end{figure}

\begin{figure*}
\begin{center}
 {\includegraphics[width=85mm,height=68mm]{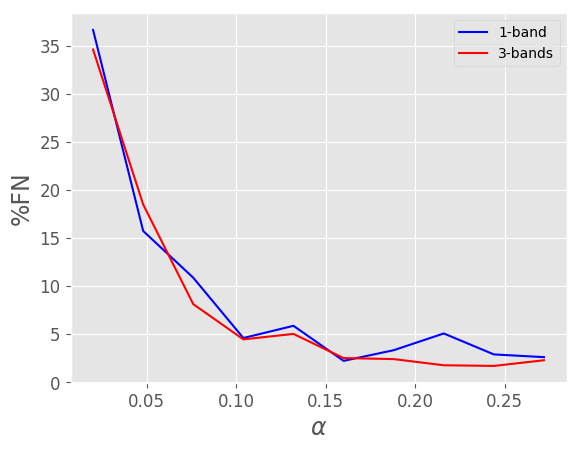}}
 {\includegraphics[width=85mm,height=68mm]{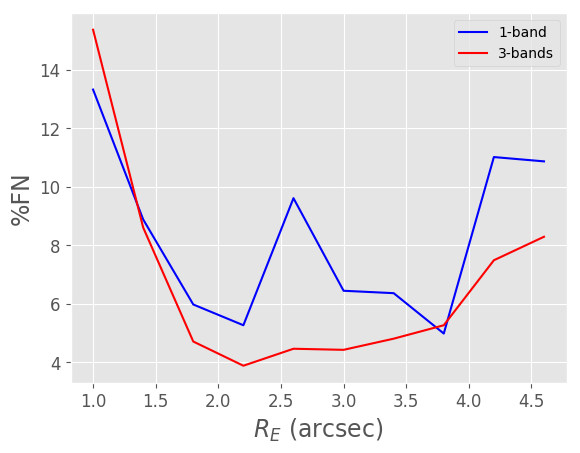}}
\caption{Percentage of false negatives (i.e., the percentage of lenses that have been misclassified) for bins of $R_E$ and $\alpha$ defined as the ratio between the peak brightness of the lens galaxy and the lensed source.}
\label{FIG_fp_er_cont}
\end{center}
\end{figure*}

\begin{figure}
\begin{center}
 {\includegraphics[width=90mm]{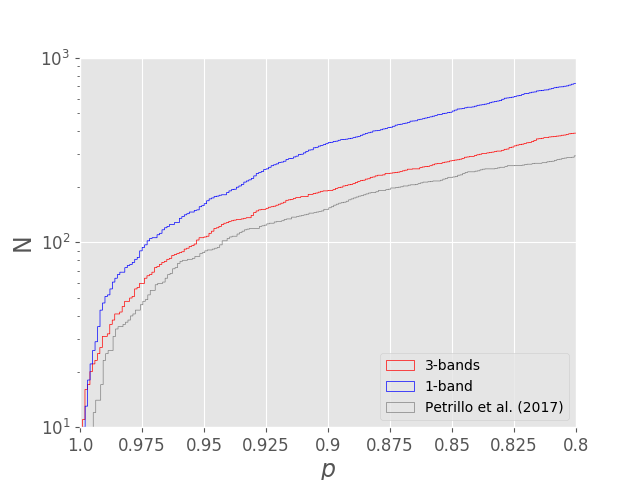}}
\caption{Number of detections as a function of the threshold for $p$ for the 1-band  (blue) and the 3-bands ConvNet (red) compared to the ConvNet of \citet{Petrillo2017}(grey).}
\label{n_detections}
\end{center}
\end{figure}

\subsection{Performance metric}

To evaluate the performances of the ConvNets we use:
\begin{itemize}
\item the true-positive rate (TPR), which measures the fraction of positive objects (in our case the lenses) detected by the algorithm. It is given by the ratio between the number of real positive (the number of real lenses that algorithm finds) and the sum of the latter and the number of false negatives (the lenses that the algorithm does not find):
 \begin{equation}
    \mathrm{TPR} = \dfrac{N_{\mathrm{TruePositives}}}{N_{\mathrm{TruePositives}}+N_{\mathrm{FalseNegatives}}} \in [0,1] ;
\end{equation}

\item the false-positive rate (FPR), which measures the fraction of negative objects (non-strong-lensing systems) misclassified as positives (lenses). It is given by the ratio between the number of false positive (the number of not lenses that algorithm misclassifies as lenses) and the sum of the latter and the number of true negatives (the non lenses that the algorithm classifies correctly)
 \begin{equation}
    \mathrm{FPR} = \dfrac{N_{\mathrm{FalsePositives}}}{N_{\mathrm{TrueNegatives}}+N_{\mathrm{FalsePositives}}} \in [0,1] ;
\end{equation}
\item these two quantities can be used to build Receiver Operating Characteristic (ROC) curves which allow to check at a glance the degree of completeness and contamination of a binary classifier. ROC curves are created by plotting $\mathrm{TPR}$ as a function of $\mathrm{FPR}$ varying the threshold of detection for $p$ between 0 and 1. This allows us to tune the value for the threshold for $p$ in order to get the desired amount of $\mathrm{TPR}$ and $\mathrm{FPR}$ for a given classification problem. In our case $p$ is the output of the ConvNet and we can tune the $p$-threshold depending how many lens candidates we desire and what level of contamination is deemed to be acceptable.
\end{itemize}

\subsection{Performance}
The ROC curves for a test-set composed of 5000 mock strong-lensing systems and 5000 non-strong-lensing systems created, as described in Sec. 2, are shown in Fig. \ref{ROC curves}. In general the 3-bands ConvNet has a better performance than the 1-band ConvNet, retrieving more mock strong-lensing systems than the 1-band ConvNet. On the contrary, the 1-band ConvNet is less contaminated by false positives at higher values of the threshold for $p$. 
Since gravitational lenses are rare events, it is important to keep a low value of FPR. Otherwise a candidate sample selected from real data would be dominated by false positives and a large amount of time would be needed to discard them through a visual inspection.
In \Fig\ref{FIG_fp_er_cont} we show, for a fiducial value for the threshold of the detection $p=0.8$, the percentage of false negatives (i.e., the percentage of lenses that have been misclassified) as a function of the Einstein radius, $R_E$, and the source over lens-galaxy brightness contrast, $\alpha$, defined in \Sec\ref{SECcreatingtrain}. Lenses with small Einstein radii and low-contrast lensed images are, as expected, the ones with which the ConvNets struggle the most. This suggests that our mock lens training samples currently covers the range in which lenses are found most easily. Smaller lenses are effectively smeared to an unrecognizable configuration by the PSF and fainter lensed sources will be too noisy to detect. 
In \Fig\ref{FIG_fp_er_cont} we also see that the accuracy decreases for larger Einstein radii, possibly due to the fact that we covered the Einstein radius with logarithmic distribution, focusing slightly more on the small-separation systems, and secondly because their lensed images are more likely to blend in to the local environment and therefore harder to distinguish from nearby galaxies by the ConvNets.

However, because our goal is to efficiently select true strong-lenses in real astronomical observations, therefore it is necessary to assess the TPR and FPR when the ConvNets are applied to real data where the performance might be worse than on a simulated dataset.


\begin{figure*}
\begin{center}
 {\includegraphics[width=85mm]{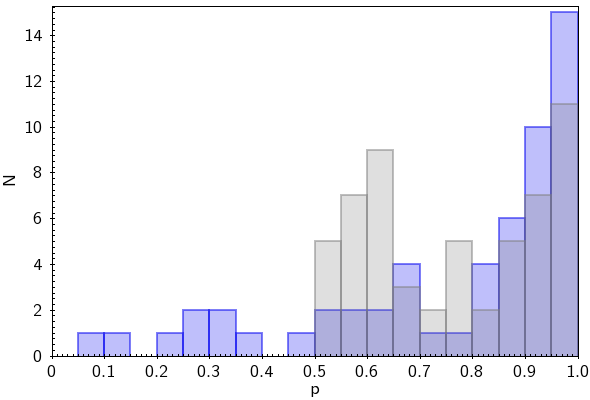}}
  {\includegraphics[width=85mm]{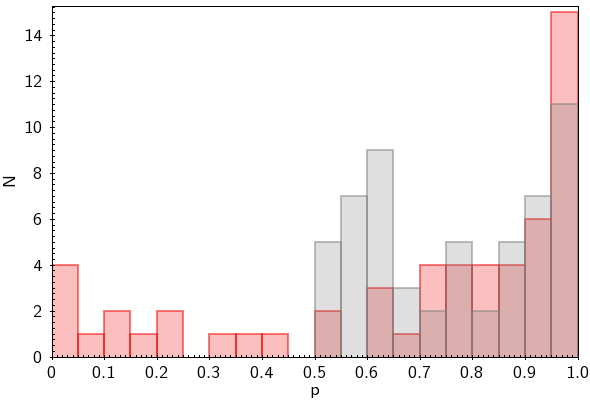}}
   {\includegraphics[width=85mm]{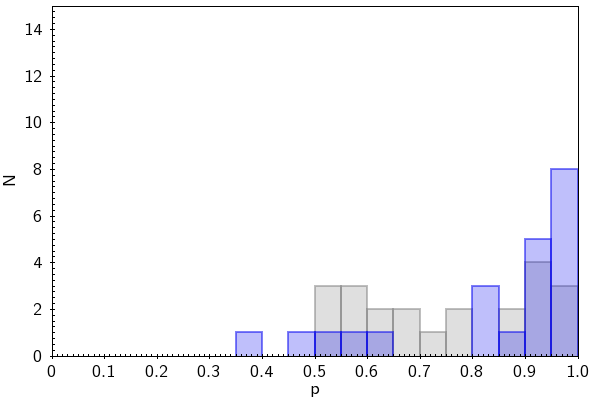}}
  {\includegraphics[width=85mm]{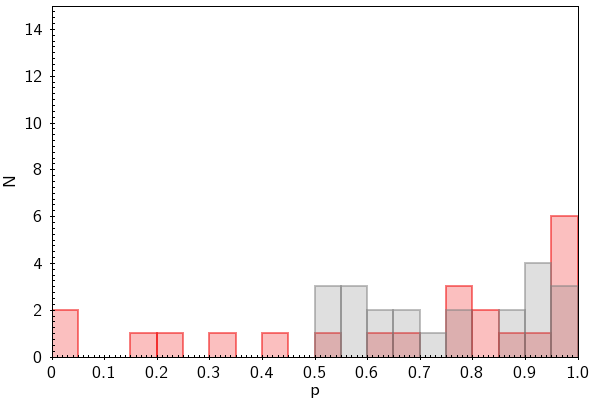}}
\caption{The first row shows the distribution of the scores for the 56 candidates with $p>0.5$ selected in \citealt{Petrillo2017}. The scores of the 1-band  (blue) and 3-bands ConvNet (red) are compared with those of \citet{Petrillo2017} (grey). The second row shows the same for a subsample of 22 candidates selected as described in \Sec\ref{SECresults}.}
\label{FIGhigh_like}
\end{center}
\end{figure*}

\begin{figure*}
\begin{center}
{\includegraphics[width=38mm]{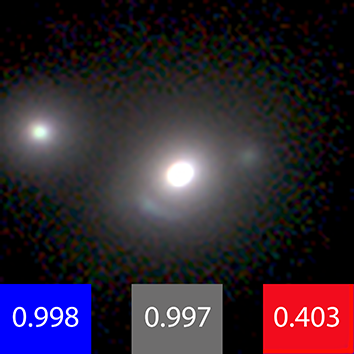}}
{\includegraphics[width=38mm]{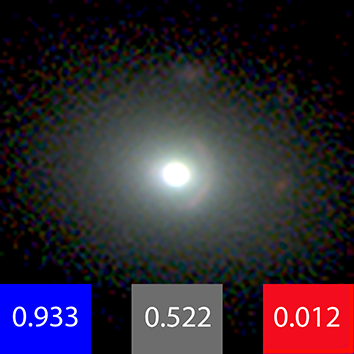}}
{\includegraphics[width=38mm]{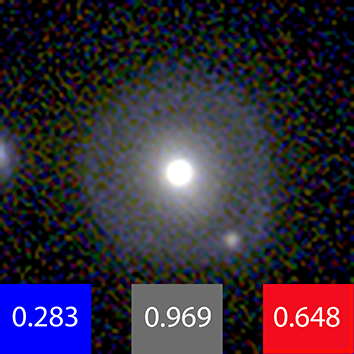}}
{\includegraphics[width=38mm]{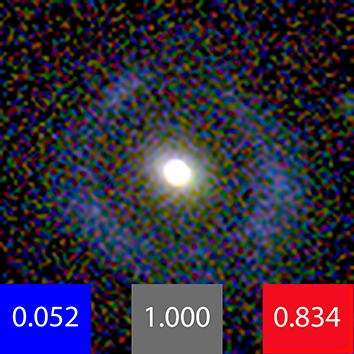}}
{\includegraphics[width=38mm]{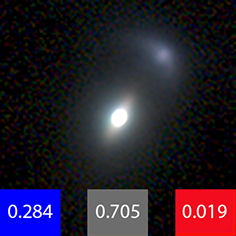}}
{\includegraphics[width=38mm]{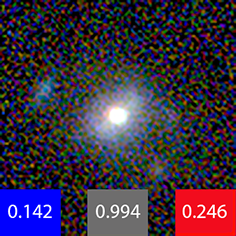}}
{\includegraphics[width=38mm]{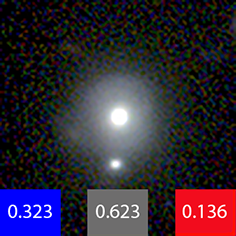}}
{\includegraphics[width=38mm]{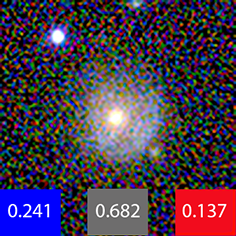}}
{\includegraphics[width=38mm]{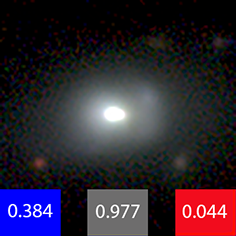}}
{\includegraphics[width=38mm]{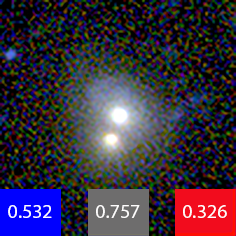}}
{\includegraphics[width=38mm]{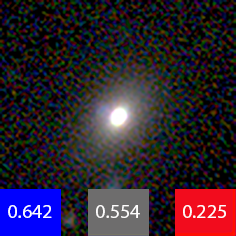}}
{\includegraphics[width=38mm]{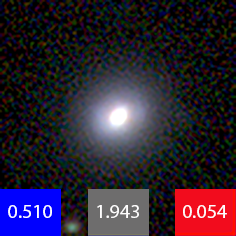}}
{\includegraphics[width=38mm]{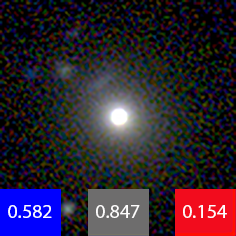}}
{\includegraphics[width=38mm]{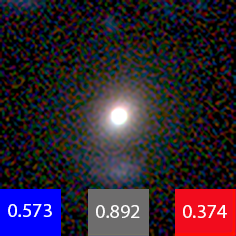}}
{\includegraphics[width=38mm]{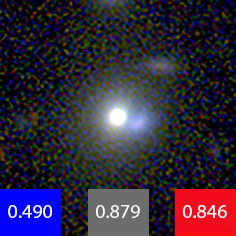}}
{\includegraphics[width=38mm]{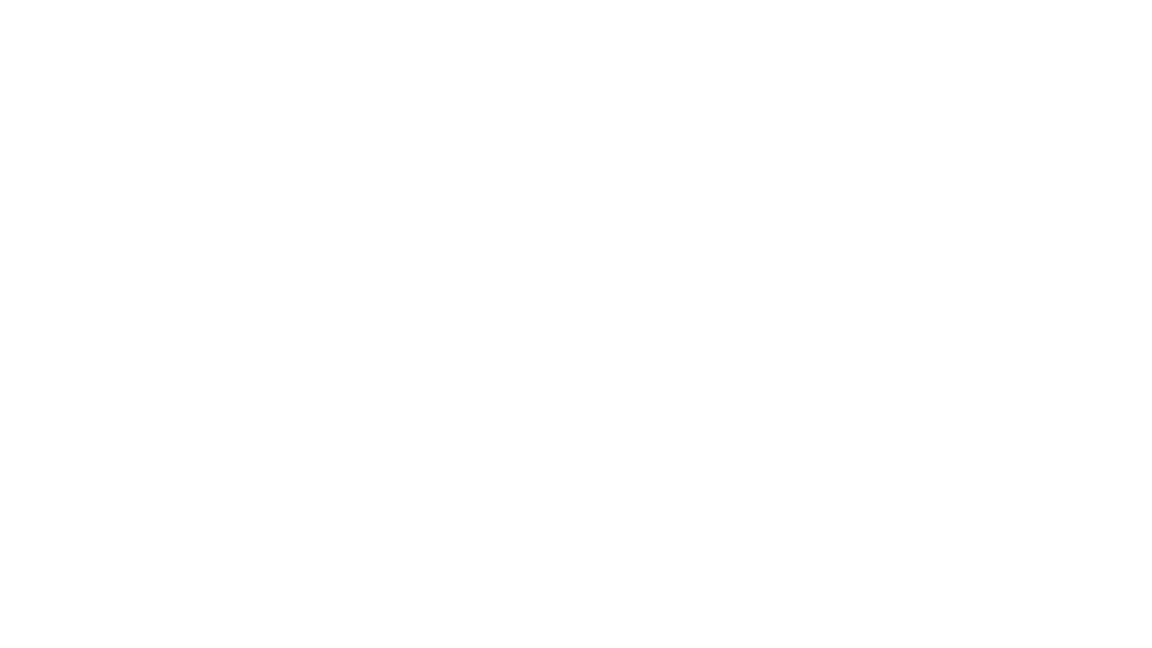}}\hspace{-2.5pt}
\caption{Sources classified by the new ConvNets with $p<0.5$ which were classified in \citealt{Petrillo2017} with $p>0.5$. On each image the score assigned by the single-band ConvNet (blue), the multi-band ConvNet (red) and \citet{Petrillo2017}'s ConvNet (grey) are reported.}
\label{FIGpm05single}
\end{center}
\end{figure*}

\section{Application to real data}\label{SECrealdata}
Testing the ConvNets on real data is fundamental, since the algorithms have been trained on a mixture of real and simulated data. It is not trivial how the method will perform on a completely real dataset, since the domain of application is slightly different with respect to the domain where the ConvNets have been trained on. 
Ideally, the morphologies in the ensemble of simulated strong-lens systems and non-strong-lens systems would be a fair representation of all morphologies observed in their equivalents in real observations.

Hence, to properly analyze the performances of the ConvNets, we apply them to the full LRG sample composed of 21789 galaxies extracted from 255 square degrees as described in \Sec\ref{sec:LRGs}.
Using the same LRG sample of \cite{Petrillo2017} allows us to assess whether there has been any improvement with respect to our previous work. 
For each galaxy image we opt to obtain an average prediction given by the average of the $p$'s for the original image and the images obtained operating a rotation of 90, 180 and 270 degrees respectively. Generally this procedure allows to increase the accuracy of the classifications.

\subsection{Results on the LRG sample}\label{SECresults}
In Fig. \ref{n_detections} we show the number of lens candidates detected varying the threshold $p$. The 1-band ConvNet detects more lens candidates compared to the 3-band one for any given threshold for $p$. 
For each of the three ConvNets it holds that the lower the threshold in p is set, the more candidates will have to be inspected visually. 
In other words, one wants to set as the threshold to an as low as possible value that yields both a sufficiently large sample of candidates and a sufficiently high TPR for the purpose of the scientific project.  
In \cite{Petrillo2017} we used visual inspection to select visually promising strong-lens candidates within the sample of systems assigned with a $p>0.5$ by the ConvNet. This sample contains 56 candidates. 
Moreover, in \cite{Petrillo2017} we selected a subsample of 22 candidates based on the agreement between their expected Einstein radii, computed from the stellar mass or the velocity dispersion of the candidate lens galaxies, and the actual galaxy-image configurations. This does not guarantee that the 22 candidates are actual lenses but it allows us to exclude the cases with more implausible configurations. \Fig\ref{FIGhigh_like} compares the $p$-values for these two samples assigned by the two new ConvNets to those assigned by \cite{Petrillo2017} ConvNet.   

We note that the $p$-values of the new ConvNets have a noticeable peak at high values that becomes even more pronounced considering only the 22 candidates. In particular, the single-band ConvNet selects high-confidence candidates assigning high values of $p$. 
This is a fair improvement of
the performance of the algorithm since there is a
larger clustering of the higher visually ranked candidates toward the high $p$-values.
Instead in \Fig\ref{FIGpm05single} we show the subset of the 56 galaxies that the new ConvNets classify with $p<0.5$.
For the 1-band finder there are not clear candidates that would be lost: maybe a couple of galaxies could be considered acceptable lens candidates, while for the rest a low $p$-value it is the ideally desired output. In particular, three candidates (third, fourth and fifth galaxy in \Fig\ref{FIGpm05single}, which have been selected as lenses in \cite{Petrillo2017} by visual inspection but after a more careful analysis have been revealed as false positives (likely a merger and two ring galaxies), are classified as non-lenses.
Thus, the new finder does cumulatively a better job in excluding contaminants and selecting lens candidates.
Instead the 3-bands lose some acceptable candidates, but more importantly misidentifies a known gravitational lens and a clear good candidate (first and second galaxy in \Fig\ref{FIGpm05single}). 
This needs further investigation, thus, in the following subsection, we analyze the behavior of the two lens-finders on a small sample composed by real lenses and striking lens candidates. 

\subsection{Application to a small sample of clear lens candidates}\label{SECrepsample}
\begin{figure*}

	\subfloat[J085446-012137]{\includegraphics[width=38mm]{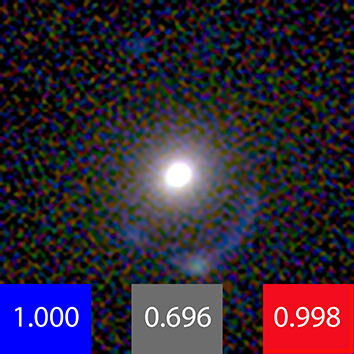}}
\hspace{\fill}
   \subfloat[J114330-014427]{\includegraphics[width=38mm]{rgb5.png}}
\hspace{\fill}
   \subfloat[J1403+0006]{\includegraphics[width=38mm]{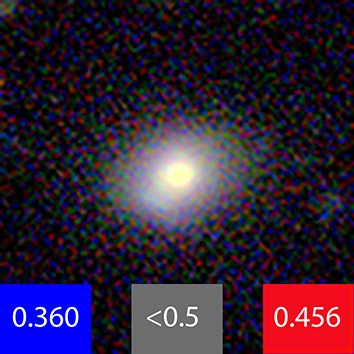}}
\hspace{\fill}
   \subfloat[J115252+004733]{\includegraphics[width=38mm]{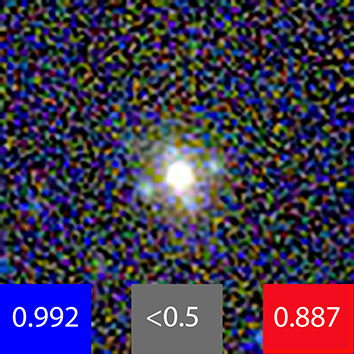}}
\hspace{\fill}
   \subfloat[KSL713]{\includegraphics[width=38mm]{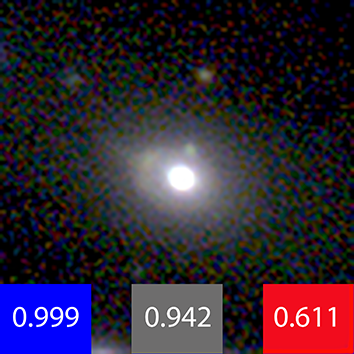}}
\hspace{\fill}
   \subfloat[KSL427]{\includegraphics[width=38mm]{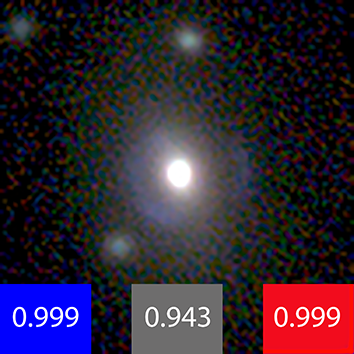}}
\hspace{\fill}
   \subfloat[KSL327]{\includegraphics[width=38mm]{KSL327.png}}
\hspace{\fill}
   \subfloat[]{\includegraphics[width=38mm]{spaziobianco.png}}
\caption{RGB images of the previously known lenses in our test-sample (first row) and the extremely likely lenses found by \citealt{Petrillo2017} in the same test-sample (second row), this sample is discussed in \Sec\ref{SECrepsample}. The scores assigned by the single-band ConvNet (blue), the multi-band ConvNet (red) and \citet{Petrillo2017}'s ConvNet (grey) are reported on each image.}
 \label{FIGrepresentative} 
 \end{figure*}

Additional insights on ConvNet performance can be obtained from inspecting the results on a set of real lenses and striking lens candidates. 
We gather a set of 6 galaxies composed as follows (see Fig. \ref{FIGrepresentative}):
\begin{itemize}
\item The four confirmed lenses known in literature which are present in our LRG sample: J085446-012137 \citep{Limousin2010}, J114330-014427 and J1403+0006 \citep{SLACS2008}, 
J115252+004733 \citep{More2017};
\item  Three extremely likely lenses found in \cite{Petrillo2017}, i.e. KSL713, KSL327 and KSL427; 
\\

The $p$-values for each of these galaxies are shown in \Fig\ref{FIGrepresentative} and \Tab\ref{TABpvalues}. It is immediately noticeable that the 1-band ConvNet is the best-performing one: the 3 extremely likely lenses and 3 of the 4 known lenses are identified correctly with a very high $p$-value; 
Instead, the 3-bands ConvNet gives more weight
to the source colours: it identifies easily blue-features but it struggles with other colours. This could be due to the larger number of training examples with blue colours. 

Moreover, both the new ConvNets presented in this paper are able to pick up the quad lens J115252+004733 not selected in \citep{Petrillo2017}. 
This piece of improvement is possibly due to the improved realism of the mock lensed sources adopted for this new ConvNets, collecting a larger variety of configurations with respect to \cite{Petrillo2017}. 
To validate this hypothesis, we train another 1-band ConvNet with the same mock lensed sources of \cite{Petrillo2017}. This new ConvNet does not detect the quad (see \Tab\ref{TABpvalues}), thus confirming the idea that the realism of the simulations plays a key-role in the performance of the lens-finder.  

To further understand the role of the simulations, we also train a ConvNet with the same configuration of \cite{Petrillo2017} but with the simulated strong-lensing systems produced for this work (see \Sec\ref{SECsims}). Also in this case the quad is detected (see \Tab\ref{TABpvalues}) even if the performance is worse than the new 1-band ConvNet. This could be due to the different architecture of the two ConvNets.

To conclude, despite the limited size of the control sample presented in this section, the 1-band ConvNet is the one which has generally the best performance under all conditions and it seems the best one to systematically apply to real data, both in terms of purity and completeness. The 3-band set-up, instead, is generally biased toward bluer colours and sometimes has failed to select known or very promising strong-lensing systems.

\end{itemize}

\begin{table*}
\caption{Scores given by ConvNets on the sample described in \Sec\ref{SECrepsample}. The 1-band and 3-band ConvNets presented in this work are compared with \citet{Petrillo2017} (old ConvNet). In the 5th column are shown the scores for a ConvNet with the same architecture of the 1-band ConvNet but trained with the mock lensed sources used in \citet{Petrillo2017} (1-band/old mocks). In the 6th column are reported the scores for a ConvNet with the same architecture of \citet{Petrillo2017} but trained with the same training set of this work (old ConvNet/new mocks).}
	\begin{center}
		\begin{tabular}{l c c c c c}
\hline
ID &  1-band & 3-bands & old ConvNet & 1-band/old mocks & old ConvNet/new mocks \\
\hline
\multicolumn{6}{c}{}\\
J085446-012137 & 1.000 & 0.998 & 0.696  & 0.148 & 0.996 \\
J114330-014427 &  0.998 & 0.403 & 0.997 & 0.947 & 0.605  \\
J1403+0006     & 0.360 & 0.456 & <0.5   & 0.038 & 0.419 \\
J115252+004733 &  0.992 &  0.887 & <0.5 & 0.321 & 0.904 \\
KSL713 		   & 0.999 & 0.611 & 0.942  & 0.990 & 0.942 \\
KSL427 		   & 0.999 & 0.999 & 0.943  & 0.982 & 0.752 \\
KSL327 		   & 0.933 & 0.012 & 0.522  & 0.587 & 0.479 \\
\hline
\end{tabular}
\label{TABpvalues}
\end{center}
\end{table*}

\captionsetup[subfigure]{labelformat=empty}
 
\begin{figure*}
\begin{center}
 {\includegraphics[width=88mm]{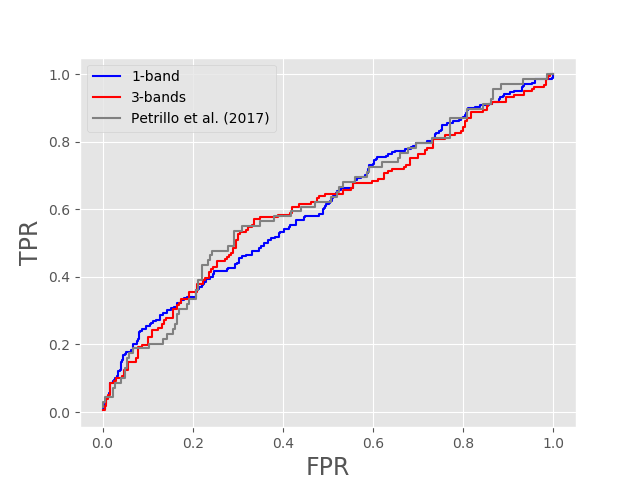}}
  {\includegraphics[width=88mm]{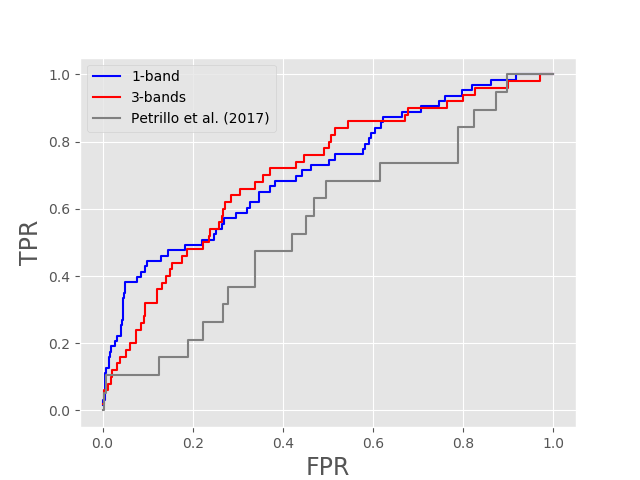}}
 {\includegraphics[width=88mm]{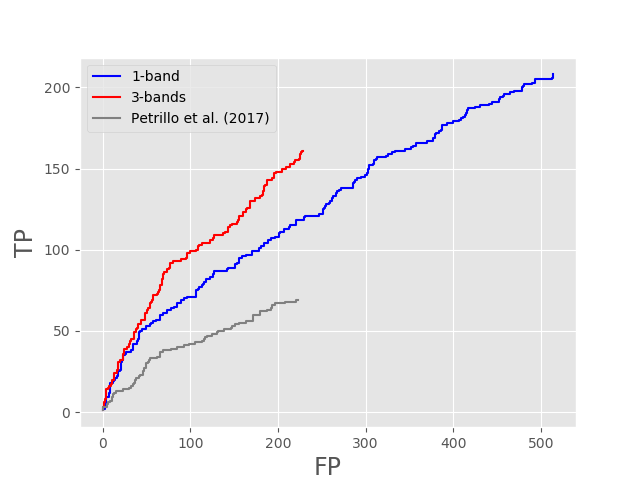}}
  {\includegraphics[width=88mm]{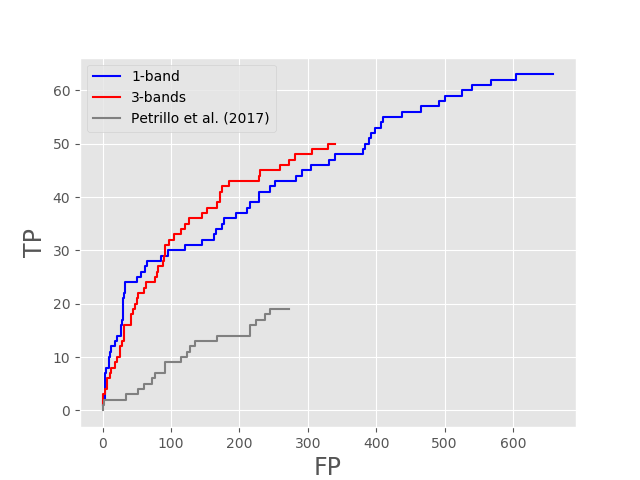}}
\caption{In the first row we show ROC curves of the 3 ConvNets built for a real sample of Luminous Red Galaxies (LRGs) as described in \Sec\ref{SECvisisp}. 
On the left is shown the result for a sample including also more dubious candidates; on the right for a sample with solid candidates only. 
On the bottom row we plot the absolute number of True Positive vs. False positives for the same samples.}
\label{FIGROCreal}
\end{center}
\end{figure*}

\subsection{Visual inspection}\label{SECvisisp}
To further analyze the performances of the ConvNets, we visually inspect the sources in the LRG sample (see \Sec\ref{SECrealdata}) selected by the ConvNets with a threshold $p>0.8$. To compare the results, we visually inspect the sources selected with the same threshold, $p>0.8$, by the ConvNet in \cite{Petrillo2017}. This yields 721 sources to inspect for the 1-band ConvNet, 390 for the 3-band ConvNet and 292 sources for the ConvNet of \cite{Petrillo2017} ConvNet. The visual inspection is carried out by 3 observers who have three possible choices: ``Yes, it is a lens", ``Maybe it is a lens", ``No, it is not a lens". For each source the classifiers visualize the \textit{g}, \textit{r} and \textit{i} fits files and a \textit{g-r-i} composed image with the software STIFF\footnote{\href{http://www.astromatic.net/software/stiff}{\tt http://www.astromatic.net/software/stiff}} \citep{Bertin2012}.

\begin{figure*}
\begin{center}
 {\includegraphics[width=88mm]{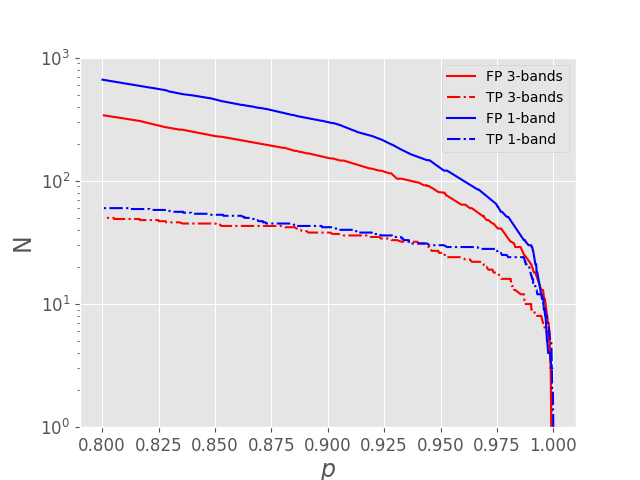}}
 {\includegraphics[width=88mm]{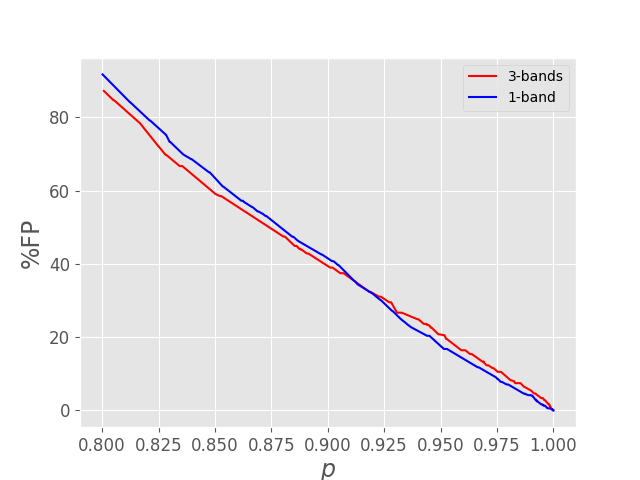}}
\caption{Results of the more conservative visual classification of the candidates selected by the ConvNets (\Sec\ref{SECvisisp}). On the left: number of False Positives and True Negatives for the 1-band and 3-bands ConvNet as a function of the threshold of detection $p$.
On the right: percentage of False Positives as a function of the threshold of detection $p$ for both the ConvNets.}
\label{FIGtp_fp_p}
\end{center}
\end{figure*}

After the visual inspection, we build ROC curves assigning a ground truth for each source from the result of the visual inspection. ROC curves are shown in \Fig\ref{FIGROCreal} for two different ground truths: one where we assign the label ``lens" to the sources that received at least two times \textit{Maybe}, and another where ``lenses" are the sources rated with 3 times \textit{Maybe} or at least one \textit{Yes} and one \textit{Maybe}. In this way we have two different ground truths: one where also dubious lens candidates are labeled as ``lenses", and one more conservative where only more solid lens candidates are labeled as ``lenses". 
In \Fig\ref{FIGROCreal} we also show the absolute number of true positives versus false positives as a function of the threshold $p$ for the same ground truths. 
The ConvNets presented in this paper have both higher purity and completeness with respect to \cite{Petrillo2017} and a larger number of lens candidates retrieved.
Moreover, the 1-band ConvNet has higher purity and completeness, especially for the more conservative sample, for sources with higher values of $p$ as also the results of \Sec\ref{SECrepsample} seems to indicate.
In addition, in \Fig\ref{FIGtp_fp_p} we show for the conservative classification the number of candidates detected by the ConvNets varying the threshold for $p$ and how many of these are False Positives.
The True Positives 
and False Positives 
do not grow at the same pace as a function of the threshold $p$ of detection. 
Moreover, the percentage of false positives is similar for the two new ConvNets. 
From the visual inspection analysis we conclude that the 1-band ConvNet would be the best choice if one wants to have a small sample of good lens candidates with few contaminants and thus less images to inspect visually. 

\subsection{Prospects for Euclid}\label{SECprospects}
We have employed about 1 hour to visually inspect the 721 targets selected by the 1-band ConvNet, selecting 60 lens candidates in the more conservative case  
(see previous subsection). For current surveys, investing such an amount of time in visual inspection is still reasonable for obtaining a more complete sample of lens candidates. However, for future surveys, the time dedicated to visually inspect the targets must be minimized considering the extremely higher number of resolved targets in the survey data.
Nevertheless, one can choose a higher threshold of detection for the ConvNet and spend considerably less time in visually inspecting the targets.
For example, if we select a threshold $p=0.9919$ for the 1-band ConvNet, which is the threshold to retrieve 3 of the 4 known lenses presented in \Sec\ref{SECrepsample}, we obtain a total of 35 sources selected by the 1-band ConvNet. These 35 sources 
are composed of 
14 positives and 21 negatives, considering the more conservative result from the visual classification. This translates to roughly $\sim40\%$ purity, assuming that our visual inspection is accurate enough. 
By using the code \textsc{LensPop} by \cite{collet2015}, we estimate that there are, in 255 square degree of KiDS, $\sim125$ lenses with a Einstein radius larger than 1 arcsecond and with our LRG colour cut selection (which corresponds to a cut in redshift of $z \lsim 0.5$, as shown in Fig. 8 in \citealt{Petrillo2017}). Hence, we can very roughly say that a candidate sample selected by the 1-band ConvNet with the fiducial threshold $p=0.9919$ is $\sim40\%$ pure and $\sim11\%$ (i.e. 14 out of 125) complete by considering a population similar to that on which we trained our ConvNets. 
To translate this result in a prediction for Euclid, assuming that ConvNets will perform at least as well on the same domain on Euclid data as they do on KiDS data, we can proceed in the following way.
\cite{collet2015} predicted that there will be $\sim170000$ discoverable lenses in Euclid. If we consider only the lenses with an Einstein radius larger than 1 arcsecond and with a redshift $z < 0.5$, the number reduces to $\sim20000$. Therefore, with the fiducial threshold of $p=0.9919$, it should be possible to select, from a sample of Euclid LRGs, $\sim2200$ good gravitational lens candidates, by visually inspecting $\sim 5500$ galaxies. 
Considering that an observer needs $\sim5$ seconds to inspect a lens candidate, that would imply that $\sim8$ hours of visual inspection would be already enough to obtain a sample of lens candidates far exceeding in size to the total number of currently known gravitational lenses. Nevertheless, this is an extremely conservative estimate, since Euclid data will have better spatial sampling (0.1 arcseconds per pixel), PSF FWHM ($\sim 0.2$ arcseconds), and thus image quality, allowing algorithms and humans to better identify lensing features. Moreover, it will be possible to train the algorithms on a wider parameter space, and thus retrieving.a larger number of lenses.
We conclude however that current ConvNets, without much adaptation, can yield already enormous numbers of lenses from Euclid data without much human effort.

\section{Discussion and Conclusions}\label{SECdiscussion}
Automatic lens finders have become a standard method for finding gravitational lens candidates in survey data. 
They are crucial to cope with the large set of an estimated $\sim10^6$ gravitational lens candidates that will be produced by upcoming surveys such as Euclid and LSST. 
Therefore, it is important to build and validate such lens-finders in order to be ready for this anticipated data-avalanche. 

It has been shown that Convolutional Neural Networks (ConvNets) are one of the most promising methods for finding lens candidates (see, e.g., \citealt{Metcalf2018}). ConvNets achieve outstanding results when the data used to train the algorithm are very similar to the data where the algorithm is intended to be applied. However, this is not necessarily the case when the domain of application is different from the domain where the algorithm is trained on. There is an active field of research investigating how to adapt the algorithms, trained on one domain to other domains (see, e.g., \citealt{csurka2017domain} for a review).
In all published cases, ConvNet lens-finders are trained using simulated mock lensed sources. Moreover, in many cases, the lens-finders are tested only on fully simulated datasets. This does not ensure that the lens-finders will perform in a satisfactory way on real survey data as shown in this paper.
It is important to conduct a thorough analysis of the performances on real data. Optimally, one would like to build a benchmark where all the lens-finders can be tested and compared against. Data from surveys such as KiDS and DES, with their identified lens candidates, could be used to build such a benchmark.

In this paper we have tested two lens-finders based on a Convolutional Neural Network (ConvNet) for selecting strong gravitational lens candidates in the Kilo-Degree Survey (KiDS). One finder just uses \textit{r}-band images while the other uses RGB images composed with \textit{g}, \textit{r}, \textit{i} images.
To train the algorithms, we have generated a large sample of simulated lensing features on top of real colour-magnitude selected galaxies from KiDS. 
Both the lens-finders are able to identify real lenses and good lens candidates in the survey. 
The performance of the two lens-finders is similar but the 3-bands finder seems to under-perform when the lensed sources do not exhibit a bluer colour. This is most likely due to the fact that the mock lensed sources in the training set have mostly blue colours. Although genuine lensed sources are usually blue, this could select against non-blue sources. One way of dealing with this issue could be populating the mock lensed sources with a wider selection of colours and, in addition, dividing the training set in different classes of colours. This would help not only the ConvNet to explicitly classify sources with different colours but could also improve the general classification given the new information provided.
In any case, the power of the single-band set-up is particularly encouraging in view of the Euclid mission data which will rely especially on the VIS band.

In addition, we have tested and compared the lens-finders with a similar one presented in \cite{Petrillo2017}. The lens-finders presented in this work have a better performance, i.e., they have both a better completeness and purity (\Sec\ref{SECvisisp}) and also they tend to classify more probable lens candidates with higher output values (see \Sec\ref{SECrepsample}). 
This is a fair improvement of the performance which implies that selecting candidates with high output values we will have a purer sample which turns out to be convenient for visual inspection and/or spectroscopic follow-ups. Indeed, the larger the future survey will become (see e.g. Euclid and LSST) the more prohibiting the visual inspection of candidates will be, hence the purity of machine learning tools will be crucial for the automatic selection of sources to be set on the queue of spectrographs for their spectroscopic confirmations.
The differences between this work and \cite{Petrillo2017} can be summarized in three points:
\begin{itemize}
\item more complex mock lensed source simulations;
\item a modified ConvNet architecture;
\item a slightly larger training set. 
\end{itemize}
These differences have contributed to the improvement of the performance, but our analysis presented in \Secs\ref{SECrepsample} indicates that the main reason is the improved mock lensed sources simulations. In this work the simulated sample is more populated with smaller sized lensed source galaxies and the sources exhibit more complexity in their structure, i.e., the presence of substructures in the lensed sources and a Gaussian Random Field perturbing the lens potential.

The ConvNet lens-finders can be tuned in terms of completeness and purity according to the specific needs of the science project. 
If one wants to have a more complete sample of lenses, a low threshold of detection can be chosen and the lower purity can be corrected by visually inspecting the candidates, something still feasible for current surveys. 
On the other hand, a purer (but less complete) sample of candidates can be obtained choosing a higher threshold. We have shown in \Sec\ref{SECprospects} that by using a high threshold for detection will be already enough to retrieve in Euclid a sample of lenses exceeding in size the total number of currently known gravitational lenses.

A series of possible improvements can be applied to the lens-finders.
As we have shown, the performance strongly depends on the composition of the training set. Hence, making the lens simulations more realistic and using real gravitational lens candidates in the training set would probably improve the quality of the classification.  
Also enlarging the training set with more KiDS galaxies would probably help, as well as adding more labels for helping the ConvNets to discriminate between different classes of galaxies (e.g., adding labels for ellipticals and spirals).
Moreover, particular care can be put in producing lens simulations where the S/N is always high enough for the lensing features to be recognizable by an observer.
Another possibility would be to specialize the ConvNets in recognizing different kinds of sources (e.g., large/small-separation systems or extended/compact-sources). This could be obtained by either training different ConvNets with specialized training sets or using a single ConvNet trained with a unique training set but with multiple labels rather than with a binary classification scheme.
Instead, on the algorithm side, a trivial improvement could be the so called \textit{ensemble averaging}, i.e., averaging the output of different ConvNets in order to possibly reduce the statistical noise of the classification. An approach experimented by, e.g., \cite{Schaefer2017} for identifying strong lens simulations.

Finally, in a forthcoming paper we will apply the algorithms to $\sim900$ square degrees of the KiDS survey starting a systematic census of strong gravitational lens candidates named \textit{``LinKS"} (Lenses in the Kilo Degree Survey).

\section*{Acknowledgements}

CEP thanks Leon Doddema, Ewout Helmich, Jelte de Jong and Kateryna Frantseva for help and support.
CEP, CT, GV, and LVEK are supported through an NWO-VICI grant (project number 639.043.308). The authors thank the anonymous referee for the insightful report. 
SC has been financially supported by a grant (project number 614.001.206) from the Netherlands Organization for Scientific Research (NWO).
GVK acknowledges financial support from the Netherlands Research School for Astronomy (NOVA) and Target. Target is supported by Samenwerkingsverband Noord Nederland, European fund for regional development, Dutch Ministry of economic affairs, Pieken in de Delta, Provinces of Groningen and Drenthe. NRN acknowledges financial support from the European Union Horizon 2020 research and innovation programme under the Marie Sklodowska-Curie grant agreement N. 721463 to the SUNDIAL ITN network.
GAMA is a joint European-Australasian project based around a spectroscopic campaign using the Anglo-Australian Telescope. The GAMA input catalogue is based on data taken from the SDSS and the UKIDSS. Complementary imaging of the GAMA regions is being obtained by a number of independent survey programmes including GALEX MIS, VST KiDS, VISTA VIKING, WISE, Herschel-ATLAS, GMRT, and ASKAP, providing UV to radio coverage. GAMA is funded by the STFC (UK), the ARC (Australia), the AAO, and the participating institutions. The GAMA website is www.gama-survey.org. 
Based on data products from observations made with ESO Telescopes at the La Silla Paranal Observatory under programme IDs 177.A-3016, 177.A-3017, and 177.A-3018, and on data products produced by Target/OmegaCEN, INAF-OACN, INAF-OAPD, and the KiDS production team, on behalf of the KiDS consortium. OmegaCEN and the KiDS production team acknowledge support by NOVA and NWO-M grants. Members of INAF-OAPD and INAF-OACN also acknowledge the support from the Department of Physics and Astronomy of the University of Padova, and of the Department of Physics of University of Federico II (Naples). 
This publication has been made possible by the
participation in the Galaxy Zoo project of more than 20.000 volunteers from around the world, with almost 2 million classifications provided. Their contributions are individually acknowledged at
http://authors.galaxyzoo.org/.
The data are generated via the Zooniverse.org platform, development of which is funded by generous support, including a Global Impact Award from Google, and by a grant from the Alfred P. Sloan Foundation.





\bibliographystyle{mnras}
\bibliography{petrillo} 



\appendix
\section{Technical details}

\begin{figure}
\begin{center}
 {\includegraphics[width=75mm]{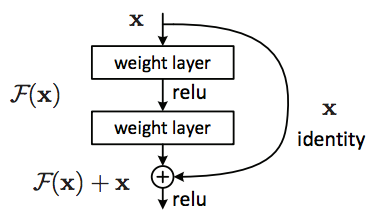}}
\caption{A building block of a ResNet. Image from \citealt{he2015deep}. \textit{Relu} is the non-linear function used between the convolutional layers; \textit{identity} is the identity function that maps the input x into itself.}
\label{FIGreslayer}
\end{center}
\end{figure}

\subsection{Convolutional Neural Networks}

The input data of a Convolutional Neural Network (ConvNet) have a topological structure, e.g., the images of cutout around galaxies that we use in this paper. 
The input images can be seen as a set of matrices $\boldsymbol{X}_k$ with $k=1,2,...,K$ 
(in the case of our 3-bands ConvNet $K=3$). 
The convolution property of the network is given by the convolutional layer 
which is composed by $n$ kernels that produce $n$ feature maps $Y_n$ as in the following equation:
\begin{equation}
\boldsymbol{Y}=\sigma\Bigg(\sum_{k=1}^K\boldsymbol{W}_k \ast \boldsymbol{X}_k+\boldsymbol{B}\Bigg),
\label{EQconvolution}
\end{equation}
where $\ast$ is the convolution operator, $\sigma$ is a non-linear function (the so-called activation function), $\boldsymbol{W}_k$ are the $K$ weight arrays with $k=1,2,...,K$, representing a filter with its bias given by the constant array $\boldsymbol{B}$. The weights $\boldsymbol{W}$ and the biases $\boldsymbol{B}$ are the parameters that are determined during the training of the algorithm. Convolutional layers are sequentially stacked such as the input of the deeper layers are the feature maps of the preceding layer.
There are far fewer parameters to be determined in a ConvNet as compared to a neural network because convolutional layers are only locally connected to its inputs as opposed to the fully connected layers of general neural networks.

It has been shown that ConvNets work better for image classification tasks than classical neural networks. One could expect this result from the following characteristics of the algorithm.
ConvNets read images preserving their structure as a 2d matrix (3d in the case of multi-band images) rather than transforming them in one-dimensional vectors where the topological information is lost.
This turns out to be important because the relevant features to classify an image (e.g, a human face) are normally localized regions of pixels (e.g., eyes, mouth, etc.) and their spatial relations (e.g., the distance between eyes and mouth).
The convolution operation of the filters with the image localizes where those patterns of pixels are in the image.
The feature maps produced with this operation highlight those predictive features in the images. 
During the training phase the optimal weights to be used in the filters are learned, which in turn means that the ConvNet learns which are the features useful for classifying the images. What usually happens is that the filters learn to extract simple features in the first layers (usually they act as edge detectors), while deeper filters learn more complicated and abstract features. See \cite{Zeiler2013} for a thorough analysis.

\subsection{Implementation}\label{architecture}
The ConvNets used in this paper are implemented in {\textsc Python 3.6} using \textsc{Keras} \citep{2015keras} with \textsc{TensorFlow} backend \citep{Tensorflow2016}.
The training phase took $\sim12$ hours on a GeForce GTX 1080. Once trained, the ConvNets take $\sim0.019$ seconds to classify one galaxy. 
For our ConvNets we use a ResNet architecture with 18 layers as described in \cite{he2015deep}. 
The building block of a ResNet (see Fig. \ref{FIGreslayer}) is characterized by a shortcut connection between the input and the output of a stack of a few convolution layers.
In this way the input $x$ is mapped into the \textit{residual} mapping $F(x)+x$. A residual mapping should be easier to optimize than an unreferenced mapping. For example in the case where an optimal mapping was the identity function, it would be easier to fit $F(x)=0$ to satisfy the requirement $F(x)+x=x$ than to satisfy the equality $F(x)=x$.

The ConvNets in this paper are trained, as in \cite{Petrillo2017}, by minimizing a loss function of the targets $t$ (1 for lenses and 0 for non-strong-lensing systems) and the predictions $p$ (the output of the sigmoid unit of the ConvNet). 
The loss function that we minimize is the binary cross-entropy, a common choice in two-class classification problems:
\begin{equation}
L = -t\log p - (1-t)\log(1-p)
\label{EQloss}
\end{equation}
The minimization is done via mini-batch stochastic gradient descent and \textit{ADAM} updates \citep{kingma2014adam} using a learning rate of 0.0001.
%
We use a batch size of 64 images and perform 80.000 gradient updates, which corresponds to about six million examples. Each batch is composed by 32 lens and 32 non-lens examples.
The weights of the ConvNets are initialized, as discussed in \cite{he2015delving}, from a random normal distribution with variance ${2/n}$ where $n$ is the number of inputs of the unit. The initial values of the biases are set to zero. 
%


\bsp	
\label{lastpage}
\end{document}